\pdfoutput=1
\documentclass[10pt,a4paper]{article}

\usepackage{amsmath,amsgen,latexsym}
\usepackage{amstext,amssymb,amsfonts,latexsym}
\usepackage{theorem}
\usepackage{pifont}
\usepackage{graphicx}
\usepackage{relsize}

 \newcommand{\bs}{\bigskip}
 \newcommand{\ms}{\medskip}
 \newcommand{\n}{\noindent}
 
 \newcommand{\hs}[1]{\hspace*{ #1 mm}}
 \newcommand{\vs}[1]{\vspace*{ #1 mm}}

 \newcommand{\setempty}{\varnothing}
 \newcommand{\real}{\mathbb{R}}
 \newcommand{\nat}{\mathbb{N}}
 
 \newcommand{\integer}{\mathbb{Z}}
 
 \newcommand{\complex}{\mathbb{C}}

 \newcommand{\AAA}{{\cal A}}

 \newcommand{\FF}{{\cal F}}

 \newcommand{\KK}{{\cal K}}

 \newcommand{\MM}{{\cal M}}
 \newcommand{\OO}{{\cal O}}
 
 \newcommand{\SSS}{{\cal S}}

 \newcommand{\matrices}[4]{\left[\:\begin{array}{cc} #1 & #2 \\%
      #3 & #4   \end{array} \right]}

\theoremstyle{plain}
\theoremheaderfont{\bfseries}
 \newtheorem{theorem}{Theorem}[section]
 \newtheorem{lemma}[theorem]{Lemma}

 {\theorembodyfont{\rmfamily}  }
 {\theorembodyfont{\rmfamily} \newtheorem{example}[theorem]{Example}}
 {\theorembodyfont{\rmfamily} }

 \newenvironment{yproof}{\par \noindent
            {\bf Proof. \hs{2}}}{\hfill$\Box$ \vspace*{3mm}}

 \newcommand{\ceilings}[1]{\lceil #1 \rceil}
 \newcommand{\floors}[1]{\lfloor #1 \rfloor}
 \newcommand{\pair}[1]{\langle #1 \rangle}

 \newcommand{\qubit}[1]{| #1 \rangle}
 \newcommand{\bra}[1]{\langle #1 |}
 
 \newcommand{\measure}[2]{\langle #1 | #2 \rangle}
 \newcommand{\density}[2]{| #1 \rangle \! \langle #2 |}

\newcommand{\ignore}[1]{}

 \newcommand{\bfzero}{{\bf 0}}

\newcommand{\ilog}{\mathrm{ilog}}

\begin{document}

\pagestyle{plain}
\pagenumbering{arabic}
\setcounter{page}{1}
\setcounter{footnote}{0}

\begin{center}
{\Large {\bf Machine Learning by Adiabatic Evolutionary Quantum System}}\footnote{A preliminary report was presented at the 22nd International Conference on Unconventional Computation and Natural Computation (UCNC 2025), Nice, France, September 1--5, 2025.} \bs\ms\\

{\sc Tomoyuki Yamakami}\footnote{Present Affiliation: Faculty of Engineering, University of Fukui, 3-9-1 Bunkyo, Fukui 910-8507, Japan}
\bs\\
\end{center}

\sloppy
\begin{abstract}
A computational model of adiabatic evolutionary quantum system (or AEQS, pronounced ``eeh-ks'') was introduced in \cite{Yam22a} as a sort of quantum annealing and its underlying input-driven Hamiltonians are generated  quantum-algorithmically by various forms of quantum automata families (including 1qqaf's). We study an efficient way to accomplish certain machine learning tasks by training these AEQSs quantumly.
When AEQSs are controlled by 1qqaf's, it suffices in essence to find an optimal 1qqaf that approximately solves a target relational problem.
For this purpose, we develop a basic idea of approximately utilizing  well-known quantum algorithms for quantum counting, quantum amplitude estimation, and quantum approximation. We then provide a rough estimation of the efficiency of our quantum learning algorithms for AEQSs.

\ms
\n{\bf keywords:} adiabatic quantum computing, quantum automata family, optimization problem, quantum counting, quantum amplitude estimation, quantum finding of maxima
\end{abstract}

\sloppy
\section{A Historical Account and an Overview of Contributions}\label{sec:introduction}

We briefly state the background of the current work and provide a quick overview of our major contributions in this work.

\subsection{Adiabatic Evolutionary Quantum Systems}\label{sec:open-problem}

Quantum physics have offered a new computing device, known as quantum computers. In theoretical fields, various models of quantum computing have been proposed, including quantum finite(-state) automata, quantum Turing machines, quantum circuits, and quantum recursion schemata \cite{Yam20}.
In many models, at each step of computation applies a unitary operation (unless making measurement). In the 1980s, Benioff \cite{Ben80} and Deutsch \cite{Deu85,Deu89} initiated a study on quantum computability based on some of these computation models.
The models that were later formulated as quantum Turing machines and quantum circuits essentially execute computations by sequentially conducting unitary  operations one by one.

Unlike such unitary-operation-based quantum computing, \emph{adiabatic quantum computing} is a physical model of performing quantum computation induced by an appropriately defined Schr\"{o}dinger equation.
A time-depending quantum state $\qubit{\psi(t)}$ evolves according to the equation $\imath \hbar  \frac{d}{dt}\qubit{\psi(t)} = H(t)\qubit{\psi(t)}$ for a specific time-dependent Hamiltonian\footnote{A Hamiltonian is a complex Hermitian matrix.} $H(t)$, where $\hbar$ is the \emph{reduced Plank constant} and $\imath$ denotes $\sqrt{-1}$.
It was proven in \cite{ADK+07} that adiabatic quantum computers are as powerful as quantum Turing machines in the polynomial-time setting.
Adiabatic computing in fact has taken advantages over solving a combinatorial optimization problem, whose goal is to find a solution that maximizes a given objective function under a certain set of constraints.
There have been numerous works demonstrating the use of adiabatic quantum computing for solving such optimization problems, e.g., \cite{FGG14}.

Lately, in \cite{Yam22a}, the computational model of so-called \emph{adiabatic evolutionary quantum systems} (or AEQS, pronounced as ``eeh-ks'') was introduced based on the Schr\"{o}dinger equation. The main target of \cite{Yam22a} was numerous  decision problems and its purpose was to determine computational resources necessary for solving the problems on AEQSs. Notice that each Schr\"{o}dinger equation is controlled by an appropriate choice of a Hamiltonian. In other words, the evolution of adiabatic quantum computation is controlled by underlying Hamiltonians.
The importance of AEQSs was demonstrated in \cite{Yam22a}, which shows that every decision problem (or equivalently, language) is solvable by an appropriate choice of AEQS with accuracy exactly $1$ (see also Theorem \ref{solvability}). This suggests an alternative way to categorize all decision problems according to the computational complexity of AEQSs that solve them.

From a practical viewpoint, it is desirable to make those Hamiltonians be ``generated'' by running families of machines. In  \cite{Yam22a}, nonetheless, those machines are taken as restricted forms of quantum automata families.
The initial goal of \cite{Yam22a} was to solve numerous decision problems (identified with languages) using AEQSs under various constraints. It has been open to develop a coherent theory on search and optimization problems. In this work, we investigate another field of machine learning and try to use  AEQSs as a new leaning model.

\subsection{Quantum Machine Learning}

We wish to seek a direct application of AEQSs to quantum machine learning.
A study of machine learning has made a crucial influence on developing artificial intelligence (AI), which is rapidly changing the society's underlying conceptual mind-set regarding intellectual activities. A widely used model for machine learning consists of various forms of neural networks. In real applications, however, a large set of parameters is required to control these neural networks for better learning consequences.

If learning a ``concept'' represented by a set of data is to find an optimal algorithm whose outcomes match the ``concept'', then our task is to search for such an algorithm approximately within a reasonable amount of time.
Thus, learning can be viewed as a process of approximating optimal solutions of some forms of optimization problems. In this work, we take this view and study efficient learning procedures in the quantum setting.

Lately, quantum machine learning has been making a crucial progress on promoting the understandings of learning process and power of quantum computing as well.
For instance, Aimeur, Brassard, and Gambs \cite{ABG07} analyzed a quantum algorithm for clustering. Moreover, Beer et. al.~\cite{BBF+20} studied a quantum analogue of neural network to accomplish given learning tasks. Our intention of this work is to provide an alternative quantum learning model to enrich our understandings of machine learning in general.
More precisely, our goal is to train an AEQS to learn a target ``concept'' by finding its underlying 1qqaf that appropriately expresses the concept.

\subsection{Major Contributions}\label{sec:contribution}

This work will provide only a set of preliminary results and leave more advanced results to an updated version of this work.
Through this work, nonetheless, we wish to demonstrate the following. Here, we focus only on simple tasks of learning certain types of ``concepts''. For simplicity, a ``concept'' is assumed to be represented as a predetermined relation (or property) on an alphabet $\Sigma$.

We wish to study how to learn a given relation (presented by a so-called \emph{supervisor}) using the model of AEQSs as accurate as possible.
As a simple, concrete example, let us consider the binary relation $EQ(x,y)$, which asserts the identity relation between two instances $x$ and $y$ of a fixed length. We must train an AEQS $\SSS$ to ``learn'' $EQ$ in the sense that $\SSS$ behaves approximately like $EQ$; in other words, $\{(x,y)\mid \SSS\text{ accepts }(x,y)\}$ is a good approximation of $\{(x,y)\mid EQ(x,y)\}$.
To simplify the basis setting further, we formulate our learning task in the following simple way. When instances are limited to a fixed length $n$, we tend to write $\SSS^{(n)}$ to denote $\SSS$ that takes only length-$n$ instances. We now restrict our attention to a simpler unary relation.

\ms
\n Simple Learning Task:
\renewcommand{\labelitemi}{$\circ$}
\begin{itemize}\vs{-2}
  \setlength{\topsep}{-2mm}%
  \setlength{\itemsep}{1mm}%
  \setlength{\parskip}{0cm}%

\item instance: a positive integer $n$.

\item target: a relation $R_n\subseteq\{0,1\}^n$, which is provided by a supervisor through queries.

\item task: train an AEQS $\SSS^{(n)}$ to learn $R_n$; that is, the set $\{x\in\{0,1\}^n \mid \SSS^{(n)} \text{ accepts }x\}$ is a ``good'' approximation of  $R_n$ with ``high'' accuracy.
\end{itemize}

Since a 1qqaf controls an AEQS, it suffices to construct a suitable 1qqaf so that the AEQS can learn a given relation properly. Hence, a key ingredient of our strategy is how to approximate a given relation $R$ by ``quantum automata,'' in particular, 1qqsf's.
An underlying idea of our approximation procedure is an application of Grover's quantum search algorithm \cite{Gro96,Gro97}.
Remember that Grover's discovery of fast quantum search algorithm has found important applications to quantum amplitude amplification \cite{BHMT02}, quantum phase/amplitude estimation \cite{Kit95}, quantum counting \cite{BHT98}, and quantum searching of minima \cite{DH96,Imr07,MIK19}.
Our learning algorithm for the above simple learning task utilizes the underlying ideas behind those quantum algorithms.

The fundamental notions and notation regarding AEQSs and 1qqaf's are provided in Section \ref{sec:fundamental}. To accomplish the above simple learning task, we make a paradigm shift from learning by AEQSs to approximating 1qqaf's in Section \ref{sec:new-setting}. The desired quantum algorithms for approximating 1qqaf's are presented in Section \ref{sec:training-AEQS}.

\section{Fundamental Notions and Notation}\label{sec:fundamental}

Let us explain the basic notions and notation necessary for the reader to read through the rest of this work. We loosely follow \cite{Yam22a} for the description of these notions and notation. For the underlying reasoning behind their introduction, the reader may refer to \cite{Yam22a}.

\subsection{Numbers, Languages, and Quantum States}

The sets of all integers, of all natural numbers (including $0$), and all real numbers are denoted $\integer$, $\nat$, and $\real$, respectively. We further set $\nat^{+}$ to be $\nat-\{0\}$. Given two integers $m$ and $n$ with $m\leq n$, the notation $[m,n]_{\integer}$ expresses the \emph{integer interval} $\{m,m+1,m+2,\ldots,n\}$. This is compared to the standard notation $[s,t]$ for the real interval $\{r\in\real\mid s\leq r\leq t\}$. We further set $[n]=[1,n]_{\integer}$ for each number $n\in\nat^{+}$. For simplicity, we define $\ilog(0)=0$ and $\ilog(x)=\ceilings{\log_2x}$ for any real number $x>0$.
Moreover, we write $\complex$ for the set of all complex numbers. We write $\imath$ for $\sqrt{-1}$. Given a number $\alpha\in\complex$, the notation $\alpha^*$ denotes its \emph{complex conjugate}. The \emph{commutator} $[A,B]$ between two square matrices $A$ and $B$ is defined to be $AB-BA$.

An \emph{alphabet} is a finite nonempty set of ``symbols'' or ``letters''. Given such an alphabet $\Sigma$, a \emph{string} over $\Sigma$ means a finite sequence of symbols taken from $\Sigma$. The notation $\Sigma^*$ denotes the set of all strings over $\Sigma$. The \emph{length} of a string $x$, denoted $|x|$, is the total number of symbols in $x$. A subset of $\Sigma^*$ is treated as a \emph{(unary) relation} over $\Sigma$ in this work whereas it is usually called a \emph{language}.
Given such a relation $R$ and an input $x\in\Sigma^*$, we often say that $R(x)$ (resp., $\neg R(x)$) is \emph{true} if $x\in R$ (resp., $x\notin R$).

In this work, we consider only quantum states in finite-dimensional Hilbert spaces. To express such quantum states, we use Dirac's ket notation.
The \emph{norm} (i.e., $\ell_2$-norm) $\|\qubit{\phi}\|$ of a quantum state $\qubit{\phi}$ is set to be $\sqrt{\measure{\phi}{\phi}}$.
A \emph{quantum bit} (or a \emph{qubit}) is a unit-norm quantum state in the Hilbert space spanned by two basis vectors $\qubit{0}$ and $\qubit{1}$ (which form the \emph{computational basis}). In contrast, the set $\{\qubit{\hat{0}},\qubit{\hat{1}}\}$ is called the \emph{Hadamard basis}, where $\qubit{\hat{0}} = \frac{1}{\sqrt{2}}(\qubit{0}+\qubit{1})$ and $\qubit{\hat{1}} = \frac{1}{\sqrt{2}}(\qubit{0}-\qubit{1})$.
For two quantum states $\qubit{\phi},\qubit{\psi}$ of the same dimension, $\qubit{\psi}$ is said to be \emph{$\varepsilon$-close} to $\qubit{\psi}$ if $\|\qubit{\phi}-\qubit{\psi}\|\leq \varepsilon$ holds.
A complex Hermitian matrix is called a \emph{Hamiltonian}. Given a Hamiltonian $H$, the \emph{spectral gap} of $H$, denoted $\Delta(H)$, is the difference between the lowest eigenvalue and the second lowest eigenvalue of $H$.
We also use the following notion of ``closeness'' between two quantum states. Given two quantum states $\qubit{\phi},\qubit{\psi}$ in the same Hilbert space, let $dis(\qubit{\phi},\qubit{\psi}) = |\measure{\phi}{\psi}|$.

For convenience, we use the notation $I$ for the identity operator acting on a Hilbert space of an ``arbitrary'' dimension. Moreover, the notation $\bfzero$ is used to express the \emph{null vector}.
The \emph{Walsh-Hadamard transform} $WH$ is defined as $\frac{1}{\sqrt{2}}\matrices{1}{1}{1}{-1}$.
For any $a,b\in\{0,1\}$, we set $CNOT(\qubit{a}\qubit{b}) = \qubit{a}\qubit{b}$ if $a=0$, and $CNOT(\qubit{a}\qubit{b})=\qubit{a}\qubit{\overline{b}}$ otherwise, where $\overline{b}=1-b$. It is well-known that the set of $CNOT$ and all single-qubit gates forms a universal gate set
(see, e.g., \cite{Gru00,NC00}).
For later use, we expand the scope of $CNOT$ from two qubits to the $n$ input qubits by simply defining $CNOT_{i,j}$ to express the application of $CNOT$ to the $i$th and $j$th qubits of the $n$ input qubits. In particular, when $i=j$, $CNOT_{i,j}$ equals $I$.
Each single-qubit gate can be in general expressed as
\[
e^{\imath\psi} \matrices{e^{-\imath \beta}}{0}{0}{e^{\imath \beta}} \matrices{\cos\theta}{-\sin\theta}{\sin\theta}{\cos\theta} \matrices{e^{-\imath \alpha}}{0}{0}{e^{\imath \alpha}}
\]
using four parameters $\alpha,\beta,\theta,\psi\in[0,2\pi)$, where $e^{\imath\psi}$ is a \emph{(global) phase shift}.
We write $U_{(\psi,\alpha,\theta,\beta)}$ for this matrix.
For example, if $\alpha=\beta=\psi=0$, then $U_{(\psi,\alpha,\theta,\beta)}$ simply represents the rotation at angle $\theta$.
We further expand $U_{(\psi,\alpha,\theta,\beta)}$ to handle all $n$ qubits by setting $\hat{U}^{(i)}_{(\psi,\alpha,\theta,\beta)} = I^{\otimes (i-1)} \otimes U_{(\psi,\alpha,\theta,\beta)}\otimes I^{\otimes (n-i)}$ for any number $i\in[n]$.
We assert that any quantum transform can be approximated with arbitrary precision by an appropriate quantum circuit composed only of $CNOT_{i,j}$s and $\hat{U}^{(i)}_{(\psi,\alpha,\theta,\beta)}$s.
The \emph{quantum Fourier transform} $\FF_k$ is defined as
\[
\FF_k(\qubit{z}) = \frac{1}{\sqrt{k}}\sum_{d=0}^{k-1}e^{2\pi {\imath}\cdot num(z) d/k}\qubit{d},
 \]
where $num(z)$ is the natural number whose binary representation matches a binary string $z$ (by ignoring the leading $0$s).

Given a Hamiltonian, its eigenvalue is particularly called an \emph{energy} and its  eigenvector is called an \emph{eigenstate}. The lowest energy is called the \emph{ground energy} and its associated eigenstate is called the \emph{ground state}.

\subsection{Adiabatic Evolutionary Quantum Systems or AEQSs}\label{sec:def-k-sna}

We quickly review the basic definition of \emph{adiabatic evolutionary quantum system} or AEQS (pronounced ``eeh-ks'').
Formally, an AEQS $\SSS$ is a septuple $(m,\Sigma,\varepsilon, \{H^{(x)}_{ini}\}_{x\in\Sigma^*}, \{H^{(x)}_{fin}\}_{x\in\Sigma^*}, \{S^{(n)}_{acc}\}_{n\in\nat}, \{S^{(n)}_{rej}\}_{n\in\nat})$, where $m:\Sigma^*\to\nat$ is a size function (called the \emph{system size}), $\Sigma$ is an (input) alphabet, $\varepsilon$ is an \emph{accuracy bound} in $[0,1)$, both $H^{(x)}_{ini}$ and $H^{(x)}_{fin}$ are Hamiltonians acting on the Hilbert space of $2^{m(x)}$ dimension (where this space is called the system's \emph{evolution space}), and  both $S^{(n)}_{acc}$ and $S^{(n)}_{rej}$ are subsets of $\{0,1\}^n$.
The pair $(S^{(n)}_{acc},S^{(n)}_{rej})$ is collectively called an \emph{acceptance/rejection criteria pair}.
A typical example of $m$ is given as $m(x)=|x|$ for all strings $x\in\Sigma^*$.
We always demand that $H^{(x)}_{ini}$ and $H^{(x)}_{fin}$ should have \emph{unique} ground states.

The Hilbert spaces spanned by the basis vectors in $\{\qubit{u}\mid u\in S^{(n)}_{acc}\}$ and those in $\{\qubit{u}\mid u\in S^{(n)}_{acc}\}$  are respectively denoted $QS^{(n)}_{acc}$ and $QS^{(n)}_{rej}$ and also called the \emph{accepting space} and the \emph{rejecting space}.
Intuitively, the AEQS $\SSS$ is said to \emph{accept} (resp., \emph{reject})  $x$ if the ground state of $H^{(x)}_{fin}$ is ``sufficiently'' close to a certain normalized vector in $QS^{(m(x))}_{acc}$ (resp., $QS^{(m(x))}_{rej}$).

Given an input $x\in\Sigma^*$, $T_x$ denotes the minimum evolution time of $\SSS$ on $x$. By setting $H^{(x)}(t)=(1-\frac{t}{T_x})H^{(x)}_{ini} + \frac{t}{T_x}H^{(x)}_{fin}$, $\SSS$ is thought to run in time $T_x\cdot \max_{t\in[0,T_x]} \|H^{(x)}(t)\|$. The ground state of $H^{(x)}(t)$ is expressed as $\qubit{\psi^{(x)}_g(t)}$. In particular, $\qubit{\psi^{(x)}_g(0)}$ and $\qubit{\psi^{(x)}_g(T_x)}$ respectively match the ground states of $H^{(x)}_{ini}$ and $H^{(x)}_{fin}$.

A lower bound of $T_x$ is given by the \emph{adiabatic theorem} \cite{Kat51,Mes58}. For any two constants $\varepsilon,\delta>0$, if
\[
T_x\geq \Omega(\frac{\|H^{(x)}_{fin} - H^{(x)}_{ini}\|}{\varepsilon^{\delta}\min_{t\in[0,T_x]} \{\Delta(H^{(x)}(t))^{2+\delta}\}}),
 \]
then $\qubit{\psi(T_x)}$ is $\varepsilon$-close to the ground state $\qubit{\psi_g(T_x)}$ of $H^{(x)}_{fin}$, as stated in \cite{ADK+07}.

When we need to restrict our interest of $\SSS$ within inputs of length $n$, we use the special notation $\SSS^{(n)}$ in place of $\SSS$.
Yamakami \cite{Yam22a} formulated the acceptance/rejection criteria of $\SSS$ as follows. Given a relation $R\subseteq \{0,1\}^*$ and a constant $\eta\in(1/2,1]$, $\SSS$ \emph{solves} (or \emph{computes}) $R$ \emph{with accuracy at least $\eta$} if, for any $x\in\Sigma^*$, (1) there exists two unique ground states $\qubit{\psi_g^{(x)}(0)}$ of $H^{(x)}_{ini}$ and $\qubit{\psi_g^{(x)}(T_x)}$ of $H^{(x)}_{fin}$,
(2) if $R(x)$ is true, then $\qubit{\psi_g^{(x)}(T_x)}$ is $\sqrt{2(1-\eta)}$-close to a normalized quantum state in $QS_{acc}^{(m(x))}$, and (3) if $\neg R(x)$ is true, then $\qubit{\psi_g^{(x)}(T_x)}$ is $\sqrt{2(1-\eta)}$-close to a normalized quantum state in $QS_{rej}^{(m(x))}$.

The computational power of AEQSs is exemplified in the following theorem of \cite{Yam22a}.

\begin{theorem}{\rm \cite{Yam22a}}\label{solvability}
For any decision problem (or equivalently, any language) $L$ over alphabet $\Sigma$, there exists an AEQS $\SSS$ of system size $1$ for which $\SSS$ solves $L$ with accuracy $1$.
\end{theorem}

This theorem supports the idea that AEQSs are useful tool for categorizing all decision problems in terms of the computational complexity of AEQSs that solve them.

\subsection{Quantum Quasi-Automata Families for AEQSs}

In \cite{Yam22a}, certain AEQSs are ``controlled'' by machine families and the computational complexity of AEQSs is measured in terms of the complexity of those machine families.
To accomplish the simple learning task described in Section \ref{sec:contribution}, the core part of the rest of this work is therefore to search for a suitable machine family that controls an AEQS. See, e.g., \cite{AY15} for the general information on quantum finite automata.

We use the notion of quantum (finite) automata to control the Hamiltonians of  the Schr\"{o}dinger equation.
A \emph{one-way quantum quasi-automata family} (abbreviated as 1qqaf) is a family $\MM=\{M_n\}_{n\in\nat}$ of machines $M_n$ of the form $(Q^{(n)},\Sigma,\{\rhd,\lhd\}, \{A^{(n)}_{\sigma}\}_{\sigma\in\check{\Sigma}}, \Lambda^{(n)}_0, Q^{(n)}_0)$, where $Q^{(n)}$ is a finite set of inner states, $\Sigma$ is an (input) alphabet, $\rhd,\lhd$ are endmarkers, each $A^{(n)}_{\sigma}$ is a quantum operator, $\Lambda^{(n)}_0$ is an initial mixture of the form $\sum_{u\in Q^{(n)}} \gamma_u\density{u}{u}$ for real numbers $\gamma_u\geq0$, and $Q^{(n)}_0$ is a subset of $Q^{(n)}$.
Each $A^{(n)}_{\sigma}$ is defined on the Hilbert space of linear operators acting on the configuration space spanned by the vectors in $\{\qubit{q}\mid q\in Q^{(n)}\}$.
The \emph{projective measurement} $\Pi^{(n)}_0$ is the projection onto the Hilbert space $span\{\qubit{u}\mid u\in Q^{(n)}-Q^{(n)}_0\}$.

Notice that each $A^{(n)}_{\sigma}$ has a \emph{Kraus representation}  $\KK_{\sigma}=\{K_{\sigma,j}\}_{j\in[k]}$ composed of $k$ Kraus operators (or operation elements) for a constant $k\in\nat^{+}$; namely, $A^{(n)}_{\sigma}(H)=\sum_{j=1}^{k}K_{\sigma,j} H(K_{\sigma,j})^{\dagger}$ for any linear operator $H$. Here, the number $k$ is referred to as the \emph{Kraus number} of $A^{(n)}_{\sigma}$.
We always demand that $\KK_{\sigma}$ satisfies the completeness relation: $\sum_{j=1}^{k}(K_{\sigma,j})^{\dagger}K_{\sigma,j}=I$.
We further assume that all entries of $K_{\sigma,j}$ are indexed by the elements of $Q^{(n)}\times Q^{(n)}$.

Each 1qqaf produces a series of Kraus operators.
We further expand $A^{(n)}_{\sigma}$ by setting inductively  $A^{(n)}_{z\sigma}(H) = A^{(n)}_{\sigma}(A^{(n)}_z(H))$ for $\sigma\in \Sigma\cup\{\lambda\}$ and $z\in\{\rhd,\lambda\}\Sigma^*-\{\lambda\}$.
Each 1qqaf $M_n$ is said to \emph{generate} (or \emph{produce}) a quantum operator $E^{(n)}_{\rhd x\lhd} = \Pi^{(n)}_0 A^{(n)}_{\rhd x\lhd}(\Lambda^{(n)}_0) \Pi^{(n)}_0$.

A \emph{selector} refers to a function from $\Sigma^*$ to $\nat$.  For instance, the standard selector has the form $\mu(x)=|x|$. A selector $\mu$ naturally induces the \emph{input domain} $\Delta_n=\{x\in\Sigma^*\mid \mu(x)=n\}$.
A family $\{H^{(x)}\}_{x\in\Sigma^*}$ of Hamiltonians is \emph{generated by 1qqaf's} if there are a polynomially-bounded selector $\mu$ and a 1qqaf  $\MM=\{M_n\}_{n\in\nat}$ such that, for any $x\in\Sigma^*$, $M_n$ generates $E^{(\mu(x))}_{\rhd x\lhd}$ and $E^{(\mu(x))}_{\rhd x\lhd}$ coincides with $H^{(x)}$.

We remark that the time-dependent Hamiltonian $H^{(x)}(t)$ of the Schr\"{o}dinger equation depends on each input $x$. The use of a 1qqaf $\MM=\{M_n\}_{n\in\nat}$, whose elements depend only on $n$, provides a means of controlling an AEQS on ``all'' inputs $x$ of length $m(x)$.

\begin{example}\cite{Yam22a}
Let $Equal$ denote the language $\{w\mid \#_0(w) = \#_1(w)\}$ over the alphabet $\Sigma=\{a,b\}$, where $\#_b(w)$ is the total number of occurrences of $b$ in $w$. It is known that no one-way finite automata solve $Equal$. However, this language $Equal$ is solved by an AEQS controlled by a certain 1moqqaf, where  1moqqaf refers to ``measure-once'' 1qqaf (which corresponds to the case that the Kraus number is $1$).
\end{example}

\section{A New Setting for Machine Learning by AEQSs}\label{sec:new-setting}

To complete the simple learning task described in Section \ref{sec:contribution}, we use the computational model of AEQS to train it appropriately so that it learns a given relation correctly. When the AEQS is controlled by a 1qqaf, it is possible to focus on constructing such a 1qqaf. In this section, we explain how to make a paradigm shift from ``learning by AEQSs'' to ``optimizing 1qqaf's.''

\subsection{From Learning by AEQSs to Optimizing 1qqaf's}

Since each AEQS is controlled by an appropriate 1qqaf, it is possible to focus on our attention only on the approximation of 1qqaf's instead of directly designing AEQSs.
In \cite{Yam22a}, several formal languages are shown to be decidable by AEQSs with appropriate choices of underlying 1qqaf's. Those constructions actually show similar traits. By exploiting such traits, in this work, we intend to restrict the framework of AEQSs as follows.

A 1qqaf $\MM=\{M_n\}_{n\in\nat}$ consists of machines $M_n$ of the form $(Q^{(n)},\Sigma,\{\rhd,\lhd\}, \{A^{(n)}_{\sigma}\}_{\sigma\in\check{\Sigma}},
\Lambda^{(n)}_0, Q^{(n)}_0)$.
To simplify the description of the quantum learning algorithms in Section \ref{sec:training-AEQS}, we intend to consider a restricted case by fixing some of the parameters of $M_n$.
We fix a designated element, say, $q_0$ in $Q^{(n)}$ and then set $Q^{(n)}_{-} = Q^{(n)} - \{q_0\}$. Moreover, we set  $\Lambda^{(n)}_0 = \sum_{u\in Q^{(n)}_{-}} \density{u}{u}$. For each $x\in\Sigma^*$, we define $m(x)=\ilog|Q^{(\mu(x))}|$ and informally treat $Q^{(\mu(x))}$ as $\{0,1\}^{m(x)}$.
For each symbol $\sigma\in\check{\Sigma}$, $A^{(n)}_{\sigma}$ is a square matrix indexed by $Q^{(n)}\times Q^{(n)}$ and we focus on the case where the Kraus number of $A^{(n)}_{\sigma}$ is $1$. In other words,  $A^{(n)}_{\sigma}(H)$ has the form $U^{(n)}_{\sigma} H (U^{(n)}_{\sigma})^{\dagger}$ for an appropriate unitary operator  $U^{(n)}_{\sigma}$, where $H$ is an arbitrary linear operator.  In analogy to the expansion of $A^{(n)}_{\sigma}$ to $A^{(n)}_{z}$ for any string $z$, we naturally expand $U^{(n)}_{\sigma}$ to $U^{(n)}_z$. Notice that $E^{(n)}_{\rhd x\lhd}$ expresses $\Pi^{(n)}_{0} A^{(n)}_{\rhd x\lhd}(\Lambda^{(n)}_{0})\Pi^{(n)}_{0}$.
Furthermore, we set $Q^{(n)}_{0}=\setempty$ for all $n\in\nat$. This makes $\Pi^{(n)}_0$ be identified with $I$.

Next, we consider an AEQS $\SSS$ of the form $(m,\Sigma,\varepsilon, \{H^{(x)}_{ini}\}_{x\in\Sigma^*}, \{H^{(x)}_{fin}\}_{x\in\Sigma^*}, \{S^{(n)}_{acc}\}_{n\in\nat}, \{S^{(n)}_{rej}\}_{n\in\nat})$. Here, we
force $S^{(m(x))}_{acc},S^{(m(x))}_{rej}\subseteq Q^{(\mu(x))}$ for all $x\in\Sigma^*$.
We further define $H^{(x)}_{ini}$ to be $WH^{\otimes m(x)}\Lambda^{(\mu(x))}_0 (WH^{\otimes m(x)})^{\dagger}$ and set $H^{(x)}_{fin} = E^{(\mu(x))}_{\rhd x\lhd}$.

\begin{lemma}
For any $x\in\Sigma^*$, the ground state $\qubit{\phi_x}$ of $H^{(x)}_{fin}$ is of the form $U^{(\mu(x))}_{\rhd x\lhd} \qubit{q_0}$.
\end{lemma}

\begin{yproof}
Fix $x\in\Sigma^*$ and set $\qubit{\phi_x} = U^{(\mu(x))}_{\rhd x\lhd}\qubit{q_0}$. Let us calculate $H^{(x)}_{fin}\qubit{\phi_x}$ to see that $\qubit{\phi_x}$ is indeed the ground state of $H^{(x)}_{fin}$.
Since $\Pi^{(n)}_{0} = I$ for any $n\in\nat$, it follows that $H^{(x)}_{fin}\qubit{\phi_x} = U^{(\mu(x))}_{\rhd x\lhd} \Lambda^{(\mu(x))}_{0}(U^{(\mu(x))}_{\rhd x\lhd})^{\dagger} \qubit{\phi_x} = U^{(\mu(x))}_{\rhd x\lhd} \Lambda^{(\mu(x))}_{0}(U^{(\mu(x))}_{\rhd x\lhd})^{\dagger} U^{(\mu(x))}_{\rhd x\lhd} \qubit{q_0}$.
The last term is further simplified to $U^{(\mu(x))}_{\rhd x\lhd}\Lambda^{(\mu(x))}_{0} \qubit{q_0}$. Since $\Lambda^{(\mu(x))}_{0}=\sum_{u\in Q^{(\mu(x))}_{-}} \density{u}{u}$, we instantly obtain $\Lambda^{(\mu(x))}_{0}\qubit{q_0}=0$. Therefore, $H^{(x)}_{fin}\qubit{\phi_x}=0$ follows. This indicates that $\qubit{\phi_x}$ is an eigenvector of $H^{(x)}_{fin}$ and its eigenvalue is $0$.
\end{yproof}

The outcome of $\SSS$ on input $x$ depends on whether its ground state $\qubit{\phi_x}$ is close enough in distance to $QS^{(m(x))}_{acc}$ or $QS^{(m(x))}_{rej}$. As shown below, this is determined by simply making a measurement of $\qubit{\phi_x}$ in the basis of $\{\qubit{u}\mid u\in S^{(m(x))}_{acc}\}$ or $\{\qubit{u}\mid u\in S^{(m(x))}_{rej}\}$. We use the notation $(S^{(m(x))}_{acc})^{\bot}$ to denote the set $Q^{(\mu(x))}-S^{(m(x))}_{acc}$.

\begin{lemma}
Let $\qubit{\phi_x}$ denote the ground state of $H^{(x)}_{fin}$ and let $\eta\in(1/2,1]$. The following two statements are logically equivalent. (1) There exists a unit-norm quantum state $\qubit{\xi}\in QS^{(m(x))}_{acc}$ such that $\|\qubit{\phi_x}-\qubit{\xi}\|\leq \sqrt{2(1-\eta)}$. (2) After a  measurement of $\qubit{\phi_x}$ in the basis of $\{\qubit{u}\mid u\in S^{(m(x))}_{acc}\}$, we observe $u$ in $S^{(m(x))}_{acc}$ with probability at least $\eta$. The same also holds for $S^{(m(x))}_{rej}$.
\end{lemma}

\begin{yproof}
(1) $\Rightarrow$ (2). We express the ground state $\qubit{\phi_x}$ as $\qubit{\psi_1}+\qubit{\psi_2}$, where $\qubit{\psi_1}=\sum_{v\in S^{(m(x))}_{acc}} \alpha_v \qubit{v}$ and $\qubit{\psi_1}=\sum_{v\in (S^{(m(x))}_{acc})^{\bot}} \beta_v \qubit{v}$ with $\alpha_v,\beta_v\in\complex$. We define $\qubit{\tilde{\psi}_1}$ to be the normalized $\qubit{\psi_1}$, that is, $\frac{1}{\|\qubit{\psi_1}\|}\qubit{\psi_1}$. Consider the value $\|\qubit{\phi_x}-\qubit{\xi}\|$ for any unit-norm quantum state $\qubit{\xi}\in QS^{(m(x))}_{acc}$. Since this value becomes the minimum when $\qubit{\xi}=\qubit{\tilde{\psi}_1}$, by (1), it follows that  $\|\qubit{\phi_x}-\qubit{\tilde{\psi}_1}\|\leq \sqrt{2(1-\eta)}$.

After making a measurement of $\qubit{\phi_x}$ in the basis of $\{\qubit{u}\mid u\in S^{(m(x))}_{acc}\}$, we obtain $\qubit{\psi_1}$. Now, we wish to show that $\|\qubit{\psi_1}\|\geq \eta$.
We first calculate the value $\measure{\tilde{\psi}_1}{\phi_x}$. It follows that $\measure{\tilde{\psi}_1}{\phi_x}  = \bra{\tilde{\psi}_1}(\qubit{\psi_1}+\qubit{\psi_2}) = \measure{\tilde{\psi}_1}{\psi_1} = \frac{1}{\|\qubit{\psi_1}\|}\measure{\psi_1}{\psi_1} = \|\qubit{\psi_1}\|$.
Notice also that $\measure{\phi_x}{\psi_1}=\|\qubit{\psi_1}\|$. From these equalities, we obtain $\|\qubit{\phi_x}-\qubit{\tilde{\psi}_1}\|^2 = \|\qubit{\tilde{\psi}_1}\|^2+\|\qubit{\phi_x}\|^2 - (\measure{\phi_x}{\tilde{\psi}_1} + \measure{\tilde{\psi}_1}{\phi_x}) = 2-2\|\qubit{\psi_1}\|$. Thus, $2(1-\|\qubit{\psi_1}\|) \leq 2(1-\eta)$ follows. We therefore conclude that $\|\qubit{\psi_1}\|\geq \eta$, as requested.

(2) $\Rightarrow$ (1). It follows from (1) that $|\measure{\xi}{\phi_x}|^2\geq \eta$ holds for a certain unit-norm quantum state $\qubit{\xi}\in QS^{(m(x))}_{acc}$. Since $|\measure{\tilde{\psi}_1}{\phi_x}|\geq |\measure{\xi}{\phi_x}|$ holds, we obtain $|\measure{\tilde{\psi}_1}{\phi_x}|\geq \eta$. As shown before,   $\|\qubit{\phi_x}-\qubit{\tilde{\psi}_1}\|^2 = 2(1-\measure{\tilde{\psi}_1}{\phi_x})$ follows. Hence, we conclude that $\|\qubit{\phi_x}-\qubit{\tilde{\psi}_1}\| \leq \sqrt{2(1-\eta)}$. \end{yproof}

To make our later analysis simpler, we set $\mu(x)=|x|$ for all $x\in\Sigma^*$. Note that $m(x)=m(y)$ holds for all pairs $x,y\in\Sigma^*$ with $|x|=|y|$ because of $m(x)=\ilog|Q^{(\mu(x))}|$. For simplicity, we write $\bar{n}$ for $\ilog|Q^{(n)}|$.

For each machine $M_n$, we modify it into another machine $\hat{M}_n$ by changing its behavior as follows. The machine $\hat{M}_n$ starts with the initial inner state $\qubit{q_0}$ and simulates $M_n$ as reading input $x$ of length $n$. When $M_n$ halts, $\hat{M}_n$ observes its inner state. This  machine $\hat{M}_n$ is said to \emph{accept} $x$ if its observed inner state is in $S^{(\bar{n})}_{acc}$ and to \emph{reject} $x$ if the observed inner state is in $S^{(\bar{n})}_{rej}$.
We briefly call $\hat{M}_n$ an \emph{acceptor version} of $M_n$ (with respect to $\SSS$).
In what follows, we automatically set $S^{(\bar{n})}_{rej} = Q^{(n)}-S^{(\bar{n})}_{acc}$.

Although the above framework looks too restrictive, this in fact makes it possible to shift from ``learning by AEQSs'' to ``optimizing 1qqaf's.''

\subsection{Matrix-Decomposition of Quantum Automata}

Now, we shift our attention from ``learning by AEQSs'' to ``optimizing 1qqaf's.'' More precisely, we search for the best possible 1qqaf that correctly computes a given relation on most instances.

Unfortunately, there are \emph{uncountably many} possible 1qqaf's usable for AEQSs. Hence, it is necessary to restrict our attention to a ``small'' set of 1qqaf's in search of an ``optimal'' 1qqaf. To achieve this goal for a given 1qqaf $\MM=\{M_n\}_{n\in\nat}$, we hereafter focus on an acceptor version $\hat{M}_n$ of each machine $M_n$ in $\MM$.
To treat such $\hat{M}_n$ in later quantum algorithms, we require its appropriately defined encoding $\pair{\hat{M}_n}$.

Firstly, we set $\Sigma=\{0,1\}$. We demand that $Q^{(n)}\subseteq [0,n+1]_{\integer}$, and then define its encoding $\pair{Q^{(n)}}$ to be the sequence $(i_1,i_2,\ldots,i_t)$ of numbers if $Q^{(n)}=\{i_1,i_2,\ldots,i_t\}$ with $i_1<i_2<\cdots <i_k$. Since $S^{(\bar{n})}_{acc}\subseteq Q^{(n)}$, we further define  $\pair{S^{(\bar{n})}_{acc}}$ to be $(j_1,j_2,\ldots,j_s)$ if $S^{(\bar{n})}_{acc} =\{j_1,j_2,\ldots,j_s\}$ with $j_1<j_2<\cdots <j_s$.

Next, we focus on each quantum transition matrix $A^{(n)}_{\sigma}$ of $M_n$ for a symbol $\sigma\in\check{\Sigma}$.
Since $A^{(n)}_{\sigma}$ is unitary, we wish to ``decompose'' it using its quantum circuit representation given in the following way.
Let us recall $CNOT_{i,j}$ and $\hat{U}^{(i)}_{(\psi,\alpha,\theta,\beta)}$ working on $n$ input qubits.
Letting $\tau = ((i_1,\psi_1,\alpha_1,\theta_1,\beta_1), (i_2,\psi_2,\alpha_2,\theta_2,\beta_2), \ldots,(i_k,\psi_k,\alpha_k,\theta_k,\beta_k),(i,j))$,
we define $V^{(\tau)}_{\sigma}$ as $\hat{U}^{(i_1)}_{(\psi_1,\alpha_1,\theta_1,\beta_1)} \hat{U}^{(i_2)}_{(\psi_2,\alpha_2,\theta_2,\beta_2)} \cdots \hat{U}^{(i_k)}_{(\psi_k,\alpha_k,\theta_k,\beta_k)} CNOT_{i,j}$.
We call this tuple $\tau$ the \emph{design} of $V^{(\tau)}_{\sigma}$.
Since $CNOT$ and all single-qubit gate form a universal gate set, $A^{(n)}_{\sigma}$ can be expressed as $V^{(\tau_1)}_{\sigma} V^{(\tau_2)}_{\sigma} \cdots V^{(\tau_m)}_{\sigma}$ for an appropriate choice of $(\tau_1,\tau_2,\ldots,\tau_m)$.
The sequence $(\tau_1,\tau_2,\ldots,\tau_m)$ is called the \emph{(design) encoding} of $A^{(n)}_{\sigma}$ and it is denoted by $\pair{A^{(n)}_{\sigma}}$.
Finally, we define the encoding $\pair{\hat{M}_n}$ of $\hat{M}_n$ by setting it to be the sequence $(\pair{Q^{(n)}}, \pair{S^{(\bar{n})}_{acc}},  \pair{A^{(n)}_{\sigma_1}}, \pair{A^{(n)}_{\sigma_2}},\ldots, \pair{A^{(n)}_{\sigma_t}})$.

To limit the number of possible encodings, we restrict the choice of $(\tau_1,\tau_2,\ldots,\tau_m)$ by allowing only a polynomial number of tuples $(i,\psi,\alpha,\theta,\beta)$ in each $\tau_k$ (such $(\tau_1,\tau_2,\ldots,\tau_m)$'s are called ``admissible'') and thus by allowing only polynomially-growing matrix-decomposition complexity.
We say that $\hat{M}_n$ is \emph{admissible} if its encoding $\pair{\hat{M}_n}$ consists of only admissible numbers. This restriction helps us search for $\hat{M}_n$ from only a pool of exponentially many encodings instead of a pool of uncountably many possible encodings.

Here, we remark that we generally assume an efficient encoding and decoding of each series of ``numbers'' so that we can freely run $\hat{M}_n$ on any input whenever $\pair{\hat{M}_n}$ is given.

\section{How to Train AEQSs to Learn Relations}\label{sec:training-AEQS}

This section presents how to accomplish the simple learning task described in Section \ref{sec:contribution}. Meanwhile, we fix $n$ ($\in\nat^{+}$) arbitrarily and focus on a target relation $R_n$, which is simply a subset of $\{0,1\}^n$. To learn $R_n$ by AEQSs, we wish to find a good approximation of $R_n$.  Since an AEQS is essentially controlled by its underlying 1qqaf, we only deal with 1qqaf's. In the rest of this section, the notation $M_n$ generally refers to the $n$-th machine of an arbitrarily fixed 1qqaf $\MM=\{M_n\}_{n\in\nat}$ and $\hat{M}_n$ refers to an acceptor version of $M_n$. We also fix a constant $\eta\in(1/2,1]$.

To simplify the notation, as long as there is no confusion, we write ``$\hat{M}_n(x)=_{\eta} R_n(x)$'' to mean that (i) $\hat{M}_n(x)$ accepts $x$ with probability at least $\eta$ if $R_n$ is true and (ii) $\hat{M}_n(x)$ rejects $x$ with probability at least $\eta$ if $\neg R_n$ is true.

\subsection{Simple Case}\label{sec:simple-case}

We intend to present quantum algorithms to optimize $\hat{M}_n$ that solves $R_n$. As a starter, we consider a simple case where
\begin{quote}
(*) there exists an admissible machine  $\hat{M}_n$ such that $\hat{M}_n(x)=_{\eta} R_n(x)$ for all instances $x\in\{0,1\}^n$.
\end{quote}
In this simple case, our task is to find such a machine $\hat{M}_n$ (actually its encoding $\pair{\hat{M}_n}$) with the help of $R_n$. We begin with sketching the basic  quantum algorithm $\AAA$, which is a backbone of the subsequent quantum algorithm.

\bs
\n Basic Quantum Algorithm $\AAA$:
\renewcommand{\labelitemi}{$\circ$}
\begin{enumerate}\vs{-1}
  \setlength{\topsep}{-2mm}%
  \setlength{\itemsep}{1mm}%
  \setlength{\parskip}{0cm}%

\item Starting with the quantum state $\qubit{0^m}\qubit{0^n}$, generate $\frac{1}{\sqrt{s}}\sum_{m_{(n)}} \qubit{m_{(n)}}\qubit{0^n}$, where $m_{(n)}$ means the admissible encoding $\pair{\hat{M}_n}$ of a machine $M_n$ and $s$ is the total number of all possible such machines.

\item Generate $\frac{1}{\sqrt{2^n}}\sum_{x:|x|=n} \qubit{x}$ from $\qubit{0^n}$ by applying $WH^{\otimes n}$.

\item For each $m_{(n)}=\pair{M_n}$, run $M_n$ on $x$ to obtain $\hat{M}_n(x)$.

\item Query to the relation $R$ and change the qubits $\qubit{m_{(n)}}\qubit{x}$ to $-\qubit{m_{(n)}}\qubit{x}$ if $\hat{M}_n(x)=_{\eta} R_n(x)$, and $\qubit{m_{(n)}}\qubit{x}$ otherwise.
    The obtained quantum state is of the form  $\frac{1}{\sqrt{s}}\sum_{m_{(n)}}\qubit{m_{(n)}}\otimes \frac{1}{\sqrt{2^n}}\sum_{x:|x|=n}\xi_{m_{(n)},x} \qubit{x}\qubit{r_{m_{(n)},x}}$ with $\xi_{m_{(n)},x}\in\{\pm1\}$ and $r_{m_{(n)},x} \in\{0,1\}$.

\item Transform $\frac{1}{\sqrt{2^n}} \sum_{x:|x|=n}\qubit{x}$ to $\qubit{0^n}$ by applying $WH^{\otimes n}$.
\end{enumerate}

Let $\qubit{\Psi}$ denote the quantum state obtained by running $\AAA$, that is, $\qubit{\Psi}= \AAA(\qubit{0^m}\qubit{0^n})$.
We can split this quantum state $\qubit{\Psi}$ into two parts $\qubit{\Psi_0}+\qubit{\Psi_1}$, where $\qubit{\Psi_1}$ has the form $\sum_{m_{(n)}} \alpha_{m_{(n)},n}   \qubit{m_{(n)}}\qubit{0^n}$ for appropriate amplitudes  $\alpha_{m_{(n)},n}\in\complex$ and $\qubit{\Psi_0}$ has the form $\sum_{m_{(n)}}\sum_{y:|y|=n\wedge x\neq0^n} \beta_{m_{(n)},n,x} \qubit{m_{(n)}}\qubit{x}$.
From our assumption, there exists an admissible  machine $M_n$ with $m_{(n)}=\pair{\hat{M}_n}$ satisfying $\hat{M}_n(x)=_{\eta} R_n(x)$ for all $x\in\{0,1\}^n$.
It thus follows that $\sum_{m_{(n)}} |\alpha_{m_{(n)},n}|^2\geq 1/s$.

\bs
\n First Quantum Algorithm:
\renewcommand{\labelitemi}{$\circ$}
\begin{enumerate}\vs{-1}
  \setlength{\topsep}{-2mm}%
  \setlength{\itemsep}{1mm}%
  \setlength{\parskip}{0cm}%

\item Apply $\AAA$ to $\qubit{0^m}\qubit{0^n}$ and obtain $\qubit{\Psi}$. We split $\qubit{\Psi}$ into two parts $\qubit{\Psi_0}+\qubit{\Psi_1}$, as stated above.

\item Let $\zeta =\|\measure{\Psi_1}{\Psi_1}\|$ and take $\theta\in[0,\pi/2]$ satisfying $\zeta =\sin^2\theta$. In the case where $\theta$ is known, with additional qubits, there is a way to increase the success probability over $1-\varepsilon$  for a small constant $\varepsilon$ \cite{CEMM98}. Since $\theta$ is unknown in general, we first need to estimate $\theta$ by running a quantum algorithm for quantum amplitude estimation. Let $\tilde{\theta}$ denote the obtained estimation of $\theta$ and set $\tilde{\zeta} =\sin^2\tilde{\theta}$.

\item Using the value $\tilde{\zeta}$, we amplify the amplitude of $\qubit{\Psi_1}$.

\item Perform a measurement on the first register and obtain a desired machine $\hat{M}_n$ with $m_{(n)}=\pair{\hat{M}_n}$.
\end{enumerate}

In what follows, we give a series of detailed explanation of each step of the first quantum algorithm stated above.

(Step 2) In this step, we conduct quantum amplitude estimation of \cite{BHMT02} in the following way. Instead of calculating the value $\eta$, we here intend to estimate the value $\theta$.
Let the controlled-$\AAA$ operator be defined as   $\OO[\AAA](\qubit{j}\qubit{z}) = \qubit{j}\otimes \AAA^j(\qubit{z})$ for any $j\in\nat$ and $\qubit{z}=\qubit{m_{(n)}}\qubit{0^n}$.
Recall the quantum Fourier transform $\FF_k$.
We apply $(\FF_k^{-1}\otimes I) \OO[\AAA] (\FF_k\otimes I)$ to $\qubit{0}\otimes \AAA(\qubit{0^m}\qubit{0^n})$. We then obtain an approximate value $\tilde{\theta}=\frac{\pi}{k}z$, where $z$ is the value obtained by performing measurement on the first register. Finally, we obtain the desired $\tilde{\zeta}$ by calculating $\sin^2\tilde{\theta}$.

(Step 3) We then amplify the amplitude of $\qubit{\Psi_1}$ using $\tilde{\zeta}$. For this purpose, we set  $C_{\tilde{\zeta}}= \frac{2}{1-\tilde{\zeta}}\density{\Psi_0}{\Psi_0}-I$ and $C_{\AAA} = 2\density{\Psi}{\Psi}-I$, which equals  $\AAA(2\density{0^{m},0^{n}}{0^{m},0^{n}}-I)\AAA^{\dagger}$. We then define $Q= C_{\AAA} C_{\tilde{\eta}}$ and apply $Q^l$ to $\qubit{0^m}\qubit{0^n}$, where $l$ is $\floors{\pi/4\tilde{\theta}}$. We then obtain $\frac{1}{\sqrt{\tilde{\zeta}}}\sin((2l+1)\tilde{\theta}) \qubit{\Psi_1} + \frac{1}{\sqrt{1-\tilde{\zeta}}} \cos((2l+1)\tilde{\theta})\qubit{\Psi_0}$. The probability of observing $\qubit{m_{(n)}}\qubit{0^n}$ is at least $\sin^2((2l+1)\tilde{\theta})\geq 1-\tilde{\zeta}$.

\subsection{Slightly More General Case}\label{sec:general-case}

In the previous subsection, we have considered the simple case in which the aforementioned condition (*) actually holds.
Here, we study a slightly more general case where this condition (*) is not assumed. Hence, we do not use Step 5 of the basic quantum algorithm $\AAA$.

Let us consider an admissible machine $\hat{M}_n$ and set $m_{(n)}=\pair{\hat{M}_n}$. We define $r_{m_{(n)},x}$ as $r_{m_{(n)},x}=1$ if $\hat{M}_n(x)=_{\eta} R_n(x)$, and $0$ otherwise. As noted in Section \ref{sec:simple-case}, if the condition (*) holds, then it follows that $dis(\frac{1}{\sqrt{2^n}} \sum_{x:|x|=n}\qubit{x}\qubit{r_{m_{(n)},x}}, \frac{1}{\sqrt{2^n}} \sum_{x:|x|=n}\qubit{x}\qubit{1})=1$.

Since we do not assume (*) here, our goal is to find a machine $\hat{M}_n$ with $m_{(n)}=\pair{\hat{M}_n}$ such that the distance $dis(\frac{1}{\sqrt{2^n}} \sum_{x:|x|=n}\qubit{x}\qubit{r_{m_{(n)},x}}, \frac{1}{\sqrt{2^n}} \sum_{x:|x|=n}\qubit{x}\qubit{1})$ is as large as possible.
Letting $\#_{R_n}(\hat{M}_n,n)= |\{x\in\{0,1\}^n\mid r_{m_{(n)},x}=1\}|$, the distance equals $\frac{1}{2^n}\cdot |\{x\in\{0,1\}^n\mid r_{m_{(n)},x}=1\}|$, which is expressed as $\frac{1}{2^n} \cdot \#_{R_n}(\hat{M}_n,n)$. It thus suffices to compute $\#_{R_n}(\hat{M}_n,n)$.

\bs
\n Second Quantum Algorithm:
\renewcommand{\labelitemi}{$\circ$}
\begin{enumerate}\vs{-1}
  \setlength{\topsep}{-2mm}%
  \setlength{\itemsep}{1mm}%
  \setlength{\parskip}{0cm}%

\item From $\qubit{0^m}\qubit{0^n}$, generate $\qubit{\Psi} = \frac{1}{\sqrt{s}}\sum_{m_{(n)}} \qubit{m_{(n)}}\qubit{0^n}$ similarly to Step 1 of $\AAA$.

\item For each value $m_{(n)}$ with  $m_{(n)}=\pair{\hat{M}_n}$, estimate the value $\#_{R_n}(\hat{M}_n,n)$  by running a quantum counting algorithm. Let $\widetilde{\#_{R_n}}(\hat{M}_n,n)$ denote the obtained estimation of $\#_{R_n}(\hat{M}_n,n)$.

\item Assume that the obtained quantum state has the form  $\sum_{m_{(n)}} \alpha_{m_{(n)},n} \qubit{m_{(n)}}\qubit{\phi_{m_{(n)},n}}$ with $\alpha_{m_{(n)},n}\in\complex$ and $\|\qubit{\phi_{m_{(n)},n}}\|=1$. We then approximate the maximum value stored in the second register.

\item Perform a measurement on the first register.
\end{enumerate}

We explain each step of the second quantum algorithm.

(Step 2) We utilize a quantum counting algorithm of Brassard, H{\o}yer, and Tapp \cite{BHT98}. Meanwhile, we fix $m_{(n)}$ arbitrarily and start this step with the quantum state $\qubit{0^n}$.
We then transform $\qubit{0^n}$ to  $\frac{1}{\sqrt{2^n}}\sum_{x:|x|=n}\qubit{x}$ by applying $WH^{\otimes n}$. We change $\qubit{m_{(n)}}\qubit{x}$ to $-\qubit{m_{(n)}}\qubit{x}$ if $M_n(x)=_{\eta} R_n(x)$, and $\qubit{m_{(n)}}\qubit{x}$ otherwise.
For each bit $b\in\{0,1\}$, let $\qubit{\Xi_b}$ denote the quantum state  $\frac{1}{\sqrt{2^n}}\sum_{x:|x|=n\wedge r_{m_{(n)},x}=b}\qubit{x}$. Let $\gamma=\measure{\Xi_1}{\Xi_1}$.
Note that $\#_{R_n}(\hat{M}_n,n)=\gamma\cdot 2^n$.
To estimate $\#_{R_n}(\hat{M}_n,n)$, it suffices to estimate $\gamma$. The estimation of $\gamma$ can be done by running a quantum phase estimation algorithm (as stated before). Let $\tilde{\gamma}$ denote the obtained estimation of $\gamma$. From this $\tilde{\gamma}$, we obtain $\widetilde{\#_{R_n}}(\hat{M}_n,n) = \tilde{\gamma}\cdot 2^n$ with probability at least $\frac{4}{\pi^2}$.

(Step 3) For this purpose, we search for $\hat{M}_n$ with $m_{(n)}=\pair{\hat{M}_n}$ such that $\qubit{\widetilde{\#_{R_n}}(\hat{M}_n,n)}$ is the nearest value to $\qubit{2^n}$. We use a basic idea of quantum algorithm of D\"{u}rr and H{\o}yer \cite{DH96} who  demonstrated how to find the minimum value in an unstructured database.
This algorithm is founded on a quantum search algorithm of Boyer, Brassard, H{\o}yer, and Tapp \cite{BBHT98}.

We generate $\frac{1}{\sqrt{2^n}}\sum_{0\leq t<2^n}\qubit{t}$. Assume that $\qubit{\phi_{m_{(n)},n}} = \sum_{y}\beta_{m_{(n)},y}\qubit{r_{m_{(n)},y}}$. Upon a query, we change $\qubit{m_{(n)}}\qubit{r_{m_{(n)},y}}\qubit{t}$ to $-\qubit{m_{(n)}}\qubit{r_{m_{(n)},y}}\qubit{t}$ if $r_{m_{(n)},y}>t$, and $\qubit{m_{(n)}}\qubit{r_{m_{(n)},y}}\qubit{t}$ if $r_{m_{(n)},y}\leq t$. Let $\qubit{\Psi}$ denote the obtained quantum state. We then split it into two parts $\qubit{\Psi_0}+\qubit{\Psi_1}$, where $\qubit{\Psi_1}$ has the form $\frac{1}{\sqrt{2^n}} \sum_{t}\sum_{m_{(n)}}\sum_{y: r_{m_{(n)},y}>t} \gamma_{m_{(n)},y,t} \qubit{m_{(n)}}\qubit{r_{m_{(n)},y}}\qubit{t}$ and
$\qubit{\Psi_0}$ has the form $\frac{1}{\sqrt{2^n}} \sum_{t}\sum_{m_{(n)}}\sum_{y: r_{m_{(n)},y}\leq t} \gamma_{m_{(n)},y,t} \qubit{m_{(n)}}\qubit{r_{m_{(n)},y}}\qubit{t}$. After running a quantum amplitude amplification algorithm, we can observe $\qubit{m_{(n)}}\qubit{\widetilde{\#_{R_n}}(\hat{M}_n,n)}\qubit{t}$. We then replace $t$ by $\widetilde{\#_{R_n}}(\hat{M}_n,n)$ and continue the above procedure. Note that the value of $t$ increases to reach the maximum value.

It turns out that the query complexity of the entire algorithm is $O(\sqrt{2^n})$.

\section{Summary and Open Problems}

Throughout this work, we have engaged in a task of learning a ``unary''  relation, which is simply a subset of $\{0,1\}^n$ and we have attempted to
complete the simple learning task given in Section \ref{sec:contribution} using AEQSs, which were originally introduced in \cite{Yam22a}.
In particular, we have focused on AEQSs whose Hamiltonians are generated by running 1qqaf's.
Our core strategy is based on a paradigm shift from learning by AEQSs to approximating 1qqaf's.
We have then provided two quantum algorithms to approximate these 1qqaf's.

In this preliminary report, we have discussed only the simple learning task and left more general tasks untreated. It is thus desirable to study more general tasks and more complex machine families, such as 2-way quantum automata families as well as pushdown automata families.
Moreover, it may be desirable  to simplify the entire quantum algorithms by combining several procedures taken in the first and the second quantum algorithms in Section \ref{sec:training-AEQS}.

There remain a number of important questions unsolved in this work. A few of them are listed below.

\renewcommand{\labelitemi}{$\circ$}
\begin{enumerate}\vs{-1}
  \setlength{\topsep}{-2mm}%
  \setlength{\itemsep}{1mm}%
  \setlength{\parskip}{0cm}%

\item It is desirable to conduct a detailed complexity analysis of the two learning algorithms given in Section \ref{sec:training-AEQS}.

\item For each choice of quantum automata family (1qqaf's and other models), which class of relations is learnable by AEQSs?

\item Are more relations learnable efficiently by AEQSs controlled by quantum automata families than by any classical learning algorithms?
\end{enumerate}


\let\oldbibliography\thebibliography
\renewcommand{\thebibliography}[1]{%
  \oldbibliography{#1}%
  \setlength{\itemsep}{-2pt}%
}

\bibliographystyle{alpha}

\end{document}